# SILICON ON NOTHING MEMS ELECTROMECHANICAL RESONATOR

*Cédric Durand[123], Fabrice Casset[2], Pascal Ancey[1], Fabienne Judong[1], Alexandre Talbot[1], Rémi Quenouillère[2], Denis Renaud[2], Stéphan Borel[2], Brigitte Florin[2], Lionel Buchaillot[3]*

[1] STMicroelectronics, 850 Rue Jean Monnet 38926 Crolles Cedex, France
[2] CEA-LETI MINATEC, 17 Rue des Martyrs 38054 Grenoble Cedex, France
[3] IEMN/ISEN Dpt, Cité Scientifique, Av. H. Poincaré B.P. 60069 59652 Villeneuve d'Ascq, France

## ABSTRACT

The very significant growth of the wireless communication industry has spawned tremendous interest in the development of high performances radio frequencies (RF) components. Micro Electro Mechanical Systems (MEMS) are good candidates to allow reconfigurable RF functions such as filters, oscillators or antennas.

This paper will focus on the MEMS electromechanical resonators which show interesting performances to replace SAW filters or quartz reference oscillators, allowing smaller integrated functions with lower power consumption. The resonant frequency depends on the material properties, such as Young's modulus and density, and on the movable mechanical structure dimensions (beam length defined by photolithography). Thus, it is possible to obtain multi frequencies resonators on a wafer. The resonator performance (frequency, quality factor) strongly depends on the environment, like moisture or pressure, which imply the need for a vacuum package.

This paper will present first resonator mechanisms and mechanical behaviors followed by state of the art descriptions with applications and specifications overview. Then MEMS resonator developments at STMicroelectronics including FEM analysis, technological developments and characterization are detailed.

## 1. INTRODUCTION

Recent demand in single chip, multi-standard wireless transceivers has focused research efforts towards developing integrated Radio Frequency (RF) Micro or Nano Electro Mechanical Systems (MEMS or NEMS) in order to replace existing off-chip oscillators like quartz or ceramics. The small sizes of the electromechanical resonators and their single wafer, multi frequencies possibilities has focussed tremendous interest in the development of demonstrators using standard CMOS process.

*Table 1* gives an overview of the state of the art on the MEMS electromechanical resonators. We can distinguish different resonator families: vibrating beam [1], longitudinal beam [2], bulk square extensional plate [3], elliptic [4] or contour [5] mode disk, rings [6], solid dielectric capacitive gap resonator [7] and bulk mode beam [8]. We can focus on the simplest case to understand how a resonator works. For a vibrating beam, we can express the mechanical resonant frequency with the following simplified equation:

$$f_{R,Beam} = A_n \sqrt{\frac{E}{\rho}} \frac{h}{L_r^2} \quad \textit{Equation 1}$$

$A_n$ is a coefficient depending on the harmonic of vibration. E is the Young modulus, $\rho$ the density, h and $L_r$ respectively the thickness and the length of the beam which have an "out of plan" displacement (*Figure 1*). Thus, the resonant frequency depends on the material properties and on the movable mechanical structure dimensions photolithographically defined.

The vibrating beam is polarized with a DC bias $V_p$. The RF signal is applied to the resonator. At the resonant frequency, the vibrating beam presents a maximum displacement detected by an output electrode as detailed in *Figure 1*. The electrode to resonator initial gap $d_0$ has to be as small as possible because the motional resistance $R_x$ (illustrates the level of signal transmitted to the resonator) is strongly dependant on this factor as explained in the following equation [7]:

$$R_{x(airgap)} = \frac{k_r}{\omega_0 V_P^2} \frac{d_0^4}{\varepsilon_0^2 \varepsilon_r^2 S^2} \frac{1}{Q_{(airgap)}} \quad \textit{Equation 2}$$

$k_r$ is the stiffness, $\omega_0$ the pulsation, $\varepsilon_0$ the air dielectric constant, $\varepsilon_r$ the material dielectric constant, S the electrode surface, and Q the quality factor. For an integrated resonator we aim to obtain a motional resistance between 50 ohms and 10kohms.

*Cédric Durand, Fabrice Casset, Pascal Ancey, Fabienne Judong, Alexandre Talbot, Rémi Quenouillère, Denis Renaud, Stéphan Borel, Brigitte Florin, Lionel Buchaillot*
*Silicon On Nothing MEMS Electromechanical Resonator*

*Figure 1: Top view of an « out of plan » vibrating beam (schematic)*

The resonant frequency increases when the beam stiffness increases, so smaller resonator sizes induce a resonant frequency increase. Moreover, the smaller the gap, the smaller the motional resistance. Thus reducing the size is a major challenge for resonators development.

The resonator performances described in the state of the art demonstrate the electromechanical resonators potentialities for quartz replacement. However, temperature stability, packaging, motional impedance constitutes major challenges to the integration of multi frequencies MEMS or NEMS resonators in Integer Circuits (IC).

In this paper, we will first be taking a look at the applications and specifications for such components. Then we will outline our "Front End" resonator technological developments on both capacitive and Metal Oxide Semiconductor (MOS) transistor detection demonstrators. We will conclude with the IC integration perspective from a device point of view.

## 2. STATE OF THE ART ON MEMS RESONATOR AND OPPORTUNITIES

*Table 1* reveals that resonators can be classed by the vibrating direction:
- Out of plane (vertical resonators)
- In plane (lateral resonators)

Initially, people were more interested in working on vertical resonators because it was technologically easier to fabricate. The majority of recent demonstrator devices are using lateral technologies mainly because of the design flexibility: squares, disks, beams…

In terms of frequency, *Table 1* shows resonators with very different resonance frequencies that make devices hard to compare. So we consider the factor Frequency multiplied by Quality Factor (F x Q) as a comparison factor.

Compared to the quartz F.Q factor (<60MHz, Q nearly 15000, F.Q=9x10⁶), the state of the art analysis demonstrates the electromechanical resonators potentialities for quartz replacement.

*Table 1 : Electromechanical resonators state of the art*

## 3. APPLICATIONS AND SPECIFICATIONS

A simple RF transceiver architecture schematic is given in *Figure 2*. We can identify the major applications using resonators:

- Reference oscillator to provide stable and precise frequency reference. Today, quartz is commonly used.

- Voltage Controlled Oscillator (VCO) which can insure a controlled variation of the frequency with bias variation. LC resonators are used today.

- Filters to extract the useful signal from the noise. Today, Surface Acoustic Waves (SAW) or Bulk Acoustic Waves (BAW) filters are used.

The overview of the main applications and specifications is given in *Table 2*.

*Figure 2 : RF architecture view*

*Cédric Durand, Fabrice Casset, Pascal Ancey, Fabienne Judong, Alexandre Talbot, Rémi Quenouillère, Denis Renaud, Stéphan Borel, Brigitte Florin, Lionel Buchaillot*
*Silicon On Nothing MEMS Electromechanical Resonator*

| Parameter | Oscillators | VCO | Band pass Filters |
|---|---|---|---|
| Central frequency | N times 38.4MHz ($F_{0,ref}$)<br>Example : N × 2; F = 76.8MHz | 2GHz | WIMAX : 2.3 to 2.7 or 3.3 to 3.7GHz<br>WIFI : 2.4 to 2.5 or 4.9 to 5.9GHz<br>DVBH : 450 to 850 MHz<br>GSM EGSM band: 880-915 (Tx) and<br>925-960MHz (Rx)<br>GSM DSC band: 1710-1785 (Tx)<br>and 1805-1880 MHz (Rx) |
| Q factor | 100 000 / N<br>Example : N 2 ; $Q_m$=50000 | 1000 | |
| Band pass | — | — | WIMAX : 1.5 to 10MHz<br>WIFI : 20MHz<br>DVBH : 5,6,7 or 8MHz<br>GSM EGSM: 35MHz<br>GSM DSC: 75MHz |
| Phase noise | -117dBc/Hz@400kHz<br>-160dBc/Hz@20MHz | -140.7 dBc/Hz<br>@600kHz | Insertion -1.5dB (-2.5 GSM)<br>Rejection -35dB (-30 GSM) |
| Operational temperature range | -40 to +100°C | -40 to +100°C | -40 to +100°C |
| Frequency stability over temperature | +/- 0.1ppm/°C<br>(Quartz with comp.) | — | — |
| Frequency stability over aging | 10ppm (Quartz) over 10 years | — | — |
| DC Voltage | 1.2 to 5V | 2.4V | — |
| Impedance | 50Ω - 10kohm | 50Ω | — |
| Tuning range | If possible | >200-300MHz | If possible |

***Table 2** : Resonator applications and specifications*

We can make a few observations about the specifications needed for the different applications describe in Error! Reference source not found.. VCO will require frequencies as high as 2GHz with a large frequency tuning. Looking at the electromechanical resonator state of the art, we can see that the highest obtained resonant frequency is around 1.5GHz. Moreover the electromechanical resonator frequency tuning is inversely proportional to the stiffness and also the frequency. Therefore it is difficult to reach the VCO specifications. The filter applications require, in general, a large band pass whereas electromechanical resonators are characterised by a high quality factor which induces a narrow band pass. A mechanical coupling could be used to enlarge the resonator band pass [1]. But in this way, it could be difficult to reach the drastic specifications needed. So, electromechanical resonators seem to be not adapted to satisfying filtering specifications. The third main application using resonator consists in providing a stable and precise reference frequency. Today, these reference oscillators are constituted by quartz oscillators which have large size and have to be built off-chip (then reported). MEMS or NEMS electromechanical resonators are good candidates to replace quartz oscillators. Its present high frequencies compared to quartz (limited around 60MHz) and use CMOS process from an integration solution perspective.

## 4. RESONATOR DESIGN AND DETECTION

### 4.1. Analytical design & Detection

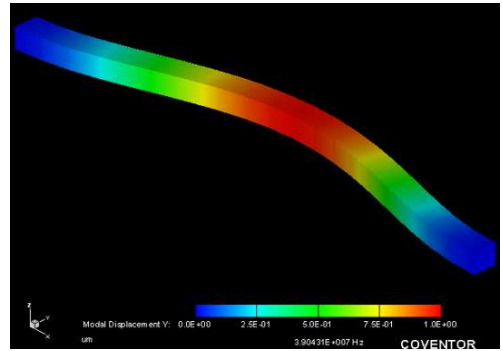

***Figure 3** : FEM modal simulation of a clamped-clamped beam*

Dimensions: Length:10µm, Width: 0.46µm, Height: 0.4µm
Coventor: 39.0MHz
Analytic: 38.8MHz
Delta: 0.5 %

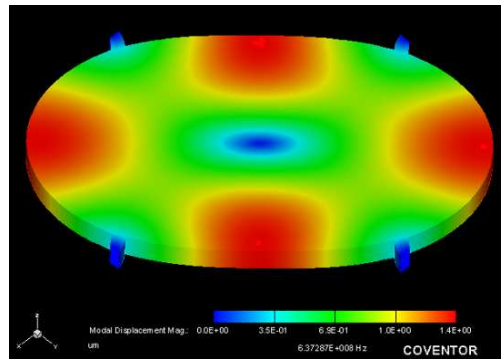

***Figure 4** : FEM modal simulation of a bulk elliptic mode disk*

Dimensions: Diameter:6µm, Height: 0.4µm
Coventor: 637MHz
Analytic: 644MHz
Delta: 1 %

We notice a good agreement between analytic and FEM results. Displacements are amplified and not representative.

It is well know that this displacement is in the nanometre range (nearly 10-20 nm for the beam, 1-5 nm for the disk). Due to very low displacements, capacitance variation is too low for output detection using capacitive detection technique. In this way, we investigate MOS detection.

## 4.2. Capacitive versus MOS Detection

MOS detection is well adapted for the output signal acquisition [9]:

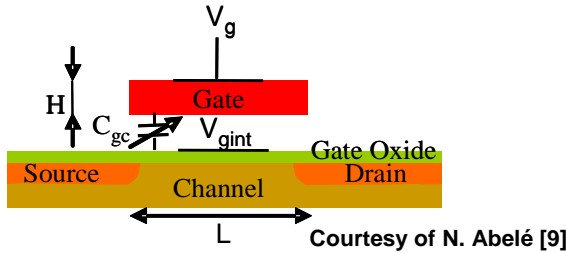

**Figure 5 :** *Schematic of a suspended resonator using MOS detection*

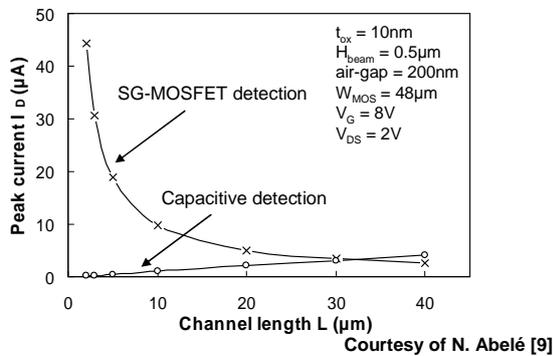

**Figure 6 :** *Comparison of MOS & Capacitive detection for a flexural mode clamped-clamped beam*

*Figure 5* gives a schematic with the principle of MOS detection. The gate is the resonant structure. *Figure 6* shows that when dimensions of resonators decrease (frequencies increase and consequently displacements decrease), the MOS detection is more efficient compared to the capacitive one. This is why MOS detection has been integrated for our technological developments.

## 5. SON TECHNOLOGICAL DEVELOPMENTS

Decrease sizes and gap is still a challenge for useful component. In this way, a front end resonator using the Silicon on Nothing (SON) process based on Silicon Germanium (SiGe) and Silicon (Si) epitaxy (*Figure 7a*) and equally used for MOS realization [10], is under development. The goal is to obtain sub 100nm gap allowing the realization of NEMS with a reduced motional resistance.

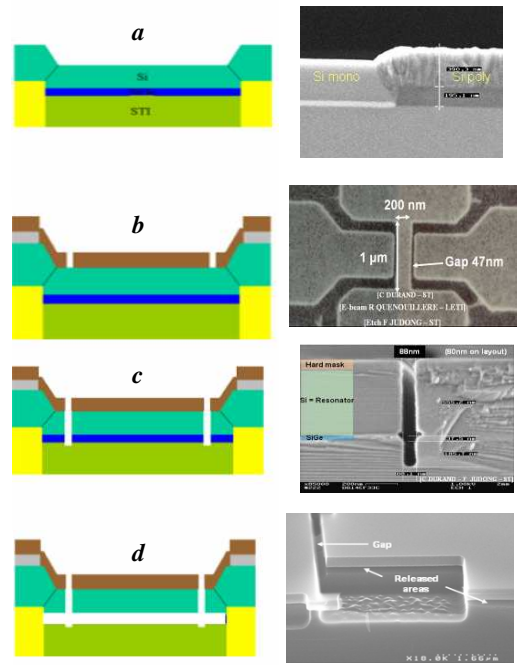

**Figure 7 :** *STMicroelectronics SON process*

Resonator gaps are made by the use of e-beam lithography (*Figure 7b*) and deep trench etching (*Figure 7c*) in order to obtain sub 100nm gaps. Finally, we etch the SiGe sacrificial layer to release the mechanical structure (*Figure 7d*). The following Scanning Electron Microscope (SEM) pictures are showing the developed technological steps.

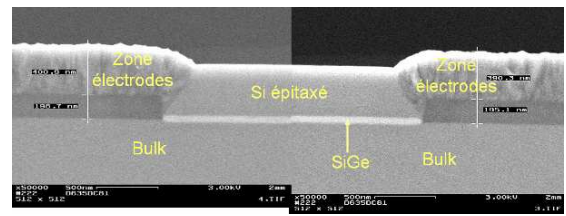

**Figure 8 :** *SEM pictures of Si & SiGe epitaxy*

*Figure 8* shows the two epitaxy steps. The first one concerns the epitaxy of the SiGe sacrificial layer. The second concerns non selective Si epitaxy that grows with theoretically unstressed monocrystalline Si above SiGe and polycrystalline Si above Silicon Oxide ($SiO_2$) layers. The resonator will be defined in monocrystalline areas by deep trench etching.

*Cédric Durand, Fabrice Casset, Pascal Ancey, Fabienne Judong, Alexandre Talbot, Rémi Quenouillère, Denis Renaud, Stéphan Borel, Brigitte Florin, Lionel Buchaillot*
*Silicon On Nothing MEMS Electromechanical Resonator*

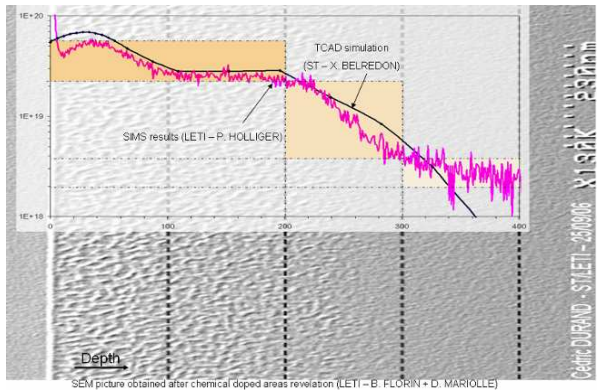

*Figure 9 : Depth doping level characterization after phosphorous doping*

To make the MOS for detection, the monocrystalline Si layer is Boron doped in situ at a level doping corresponding to the MOS channel doping, around $5.10^{15}$ at/cm$^3$. A phosphorous doping is necessary to make Source, Drain, and Gate (= resonator). *Figure 9* shows the depth doping level characterization with a comparison of simulation, SIMS method and a chemical doping level revelation [11], with a good agreement. The doping level is not continuous with depth because of technological constraints. The minimum doping level of around $3.10^{18}$ at/cm$^3$ seems to be enough for the application. The physical revelation enables the study of lateral doping diffusion (*Figure 10*).

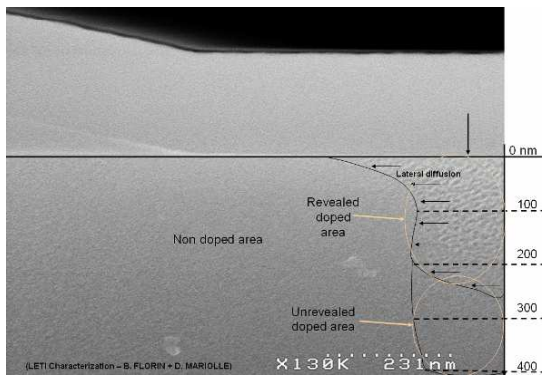

*Figure 10 : Lateral doping diffusion after thermal annealing*

In order to limit the high motional resistance of devices, it is desirable to make small gaps between resonators and electrodes. It is why the gap has been defined by e-beam photolithography and etched (*Figure 7b & c*) using a highly anisotropic plasma process. Some pictures of the technological results are presented in *Figure 11* and *Figure 12*.

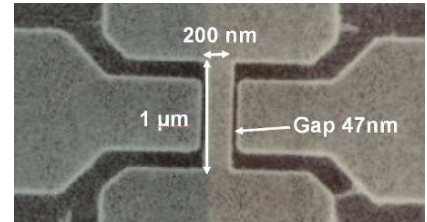

*Figure 11 : Top SEM view of a nano beam*

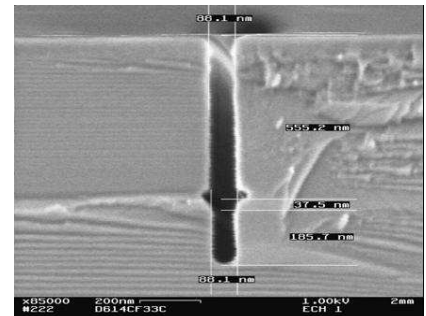

*Figure 12 : Gap etch cross SEM picture*

*Figure 11* gives an idea of the precision of technological gap definition after etching. *Figure 12* gives a cross section view a of 90nm gap obtained after a 400nm Si etching with an enlargement of less that 10nm due to the etching process.

After the gap etching, the SiGe (sacrificial layer) has to be etched (*Figure 7d*) with a high selectivity SiGe/Si and SiO$_2$/SiGe. The aim is to release resonators that will be suspended.

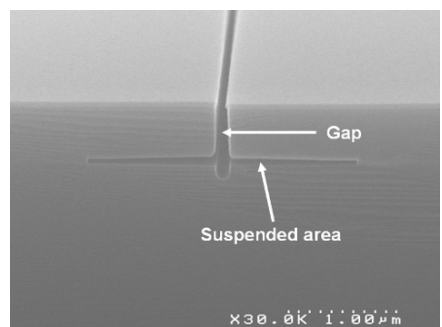

*Figure 13 : Tilt SEM view of SiGe released areas*

*Cédric Durand, Fabrice Casset, Pascal Ancey, Fabienne Judong, Alexandre Talbot, Rémi Quenouillère, Denis Renaud, Stéphan Borel, Brigitte Florin, Lionel Buchaillot*
*Silicon On Nothing MEMS Electromechanical Resonator*

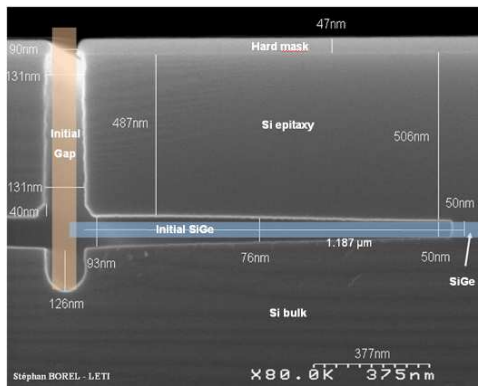

***Figure 14 :*** *Cross section view of a released tunnel*

The process has been developed initially mainly for CMOS using SON technology purposes with tunnel to etch around 500nm [12]. Here the process has been adapted to be able to etch longer tunnels as shown in **Figure 13** and **Figure 14**. The process selectivity is strongly dependant of the tunnel depth. For a 1.19µm tunnel depth, the initial gap measured at 90nm has been enlarged and measured at 130nm after release. The process is under progress so as to improve selectivity.

Work is in progress in collaboration with the CEA-LETI & IEMN to obtain electrical demonstrators in the close future.

## 6. CONCLUSION

MEMS or NEMS electromechanical resonator is a very important component in order to integrate frequency reference oscillator in chips. Some developments to obtain capacitive and MOS demonstrators were presented. The electromechanical resonator state of the art highlights the importance of sub 100nm gaps and packaging. The SON process seems to be well adapted to obtain promising components. Process developments have started from electrical demonstrators' realization. Future work on packaging and IC integration will be investigated in a close future.